\documentclass[referee]{aa}
\usepackage{graphicx}
\usepackage{amssymb}
\usepackage{natbib}
\usepackage{float}

\newcommand{\ap}[1]{\ifmmode^{\rm #1}\else $^{\rm #1}$\fi}
\newcommand{\gradi}{\ifmmode^\circ\else$^\circ$\fi}
\newcommand{\primi}{\ifmmode^{'}\else$^{'}$\fi}
\newcommand{\secon}{\ifmmode^{"}\else$^{"}$\fi}
\newcommand{\ore}{\ifmmode^h\else$^h$\fi}
\newcommand{\minu}{\ifmmode^m\else$^m$\fi}
\newcommand{\seco}{\ifmmode^s\else$^s$\fi}

\begin{document}

\title{
Formation of terrestrial planets in close binary systems:\\
the case of $\alpha$ Centauri A
}

\author{
M. Barbieri\inst{1}
\and 
F. Marzari \inst{2} 
\and 
H. Scholl  \inst{3} 
}

\authorrunning{Barbieri M., et al.}
\titlerunning{Terrestrial planet formation in $\alpha$ Centauri A}

\offprints{M. Barbieri, \email{mbarbier@pd.infn.it}}

\institute{
CISAS, c/o Dipartimento di Fisica, Universit\`a di Padova, via Marzolo, 8
I-35131 Padova, ITALY, \email{mbarbier@pd.infn.it}\\
\and
Dipartimento di Fisica, Universit\`a di Padova, via Marzolo, 8
I-35131 Padova, ITALY, \email{marzari@pd.infn.it}\\
\and
Observatoire de la Cote d'Azur, B.P. 4229, Nice Cedex 4, 
F-06304 FRANCE, \email{scholl@obs-nice.fr}\\
}

\date{Received 3 April 2002/Accepted 22 August 2002}

\abstract{
At present the possible existence of planets around the stars of a close binary 
system is still matter of debate. Can planetary bodies form in spite of the 
strong gravitational perturbations of the companion star? We study in this 
paper via numerical simulation the last stage of planetary formation, from 
embryos to terrestrial planets in the $\alpha$ Cen system, the prototype of 
close binary systems. We find that Earth class planets can grow  around 
$\alpha$ Cen A on a time-scale of 50 Myr. In some of our numerical models the 
planets form directly  in the habitable zone of the star in low eccentric orbits. 
In one simulation two of the final planets are in a 2:1 mean motion resonance 
that, however, becomes unstable after 200 Myr. 
During the formation process some 
planetary embryos fall into the stars possibly altering their metallicity.
}

\maketitle

\section{Introduction}

Planetary formation in close binary systems is still an open problem. Accretion 
disks have already been observed around each individual component \citep{rod98} 
strongly supporting the idea that at least the initial stage of planetary 
formation can occur. However, it is still uncertain  whether the dust can 
coalesce into planetesimals and, in particular, whether the process of 
planetesimal accretion can continue till the completion of a planet. \cite{msc00} 
showed that relative velocities between small planetesimals are low in spite of 
the strong gravitational perturbations due to the companion star. The combined 
effects of gas drag and secular perturbations induce a sharp periastron 
alignment of the planetesimal orbits and, as a consequence, low relative speeds 
at impact that favor accumulation rather then fragmentation. However, the 
gravitational perturbations from the companion star may still halt the planetary 
formation process in the last stage when planetary embryos collide together to 
form a planet. In this paper we numerically model this final scenario where 
large embryos collide and, eventually, form larger bodies that can be termed 
planets. 

We have concentrated on the Alpha Centauri system, the prototype of close 
binary systems with the more massive star very similar to our Sun. It is also 
the same system studied by \cite{msc00}. 
 Planetary embryos are likely to form within 
2.5 AU of the star since beyond the companion perturbations are 
too strong and
planetesimal collisions may lead preferentially to fragmentation 
rather then accretion \citep{msc00}.  Moreover, orbits farther than 
2.7 AU become unstable according to \cite{hwi99}.
We model the accumulation of Lunar--size 
embryos into terrestrial planets to estimate the time-scale of formation and 
the dynamical and physical properties of the final planets. Different initial 
conditions are considered to have a statistical description of the possible 
final systems. 

\section{The numerical model}

We have performed 26 N-body simulations where we have numerically integrated 
the orbits of planetary embryos and of the companion star for 10$^8$ 
years with the Mercury package described in \cite{cha99}. Mercury is a hybrid 
integrator, made of a conventional Burlish-Stoer integrator and a symplectic 
integrator. The symplectic integrator is preferred since it exhibits no 
long-term accumulation  of energy error, and is fast. A timestep of 2 days 
was adopted since the inner body is at 0.4 AU from the star. 

We performed some tests to verify the reliability of the symplectic algorithm 
in the presence of a massive perturber as Alpha Centauri B. Massless particles have 
been integrated for 10 Myr with semimajor axes of 2.0, 2.25, and 2.5, and 
eccentricity of 0, 0.2, 0.5 and 0.7 with both the symplectic and the Radau 
\citep{eve85} integrators. The region between 2.0 and 2.5 AU is the outermost 
zone 
initially populated by protoplanets in our simulations and the most
critical since the perturbations of the companion star are stronger. 
The outcome of the two algorithms were in good agreement and significative 
differences were found only for highly eccentric orbits ($e =0.5,0.7$) 
at 2.5 AU: 
the bodies escape on hyperbolic orbits in both the simulations, 
but at different times depending on the integrator used. 
However, this discrepancy
is not due to a bad approximation in the symplectic 
algorithm but to the chaotic behaviour of eccentric bodies close to the
2.5 AU location. To confirm this supposition, we integrated 
with Radau
two clones of the particles started at 2.5 AU and with 
eccentricity of 0.5 and 0.7, respectively. The two clones were
obtained by changing the 9th digit of the semimajor axis of 
the original test particles. We found again a significant change 
(in one case of an order of magnitude) in the escaping time even 
if the same integrator was used (Radau). As a further test we 
verified that the 
relative energy change in our simulations was of the order of $10^{-8}$. 
The Hamiltonian decomposition at 
the base of the symplectic algorithm is then still effective, at least 
within 2.5 AU from the star.
Additional tests have also been performed in the inner region of the protoplanet
disk. The numerical simulations with the two integrators showed no discrepancy,
even for large values of eccentricity, for a semimajor axis $a = 0.5$ AU,
the inner value used in our starting conditions. The choice of a short
timestep (2 days) for the symplectic algorithm was critical in this respect.

Collisions between 
embryos are assumed to be completely inelastic; fragmentation or cratering are 
not considered. This is a good approximation since the gravitational binding 
energy for large bodies overcomes the kinetic energy of the fragments also 
at high velocity impacts. A single body is left after a collision and its orbit 
is determined by the conservation of linear momentum.

The integrations were simultaneously performed in a local 
CONDOR pool (http://www.cs.wisc.edu/condor)
made of about 200 Pentium class computers with an average speed of 200 Mflops.
A single simulation needed about 2 months of real time.

\section{Initial conditions}

$\alpha$ Centauri is a triple system with two of the stars forming a close 
binary (HD 128620 and HD 128621) and a third (GJ 551) orbiting this pair at a 
much greater distance. The physical properties of the stars in the system are 
summarized in Table \ref{tab1}.

\begin{table}
\center
\begin{tabular}{cccc}
star  &    mass     &    MK class & ref.\\
      &  $M_\odot$  &             &     \\
\hline
A     &    1.1      &    G2 V     & a   \\
B     &    0.92     &    K0 V     & a   \\
C     &    0.11     &    M5 V     & b   \\
\end{tabular}
\caption{Physical properties of the stars in $\alpha$ Centauri system.
(a) \cite{pnn99}; (b) \cite{ben99}}
\label{tab1}
\end{table}

In our model, we ignore the distant third star, $\alpha$ Cen C (Proxima), as it 
might not be bound to the central binary system \citep{aop94}. Moreover, its 
perturbations on the embryo orbits would be small. 
The star $\alpha$ Cen B moves on a 
fixed elliptical orbit around the primary component A, with a semi-major axis 
of 23.4 AU and an eccentricity of 0.52 \citep{sod99}. The protoplanets 
move around component A, and they are set on the orbital plane  of the binary 
system with a low inclination with respect to it. This assumption is based on 
the work of \cite{hal94}. He shows that if the separation between 
the components of a binary system is around 30-40 AU or less, 
the components of observed binary 
systems are coplanar with respect to their mutual orbit and their equator.

To compute the initial distribution of embryos around the primary star, we have
adopted a surface density of solid material $\sigma$ equal to 8 g\,cm$^{-2}$ at 
1 AU,  the same value adopted by \cite{cha01} and intermediate between that used 
by \cite{kok96} and that adopted by \cite{cha98} for the solar nebula. Since we 
cover a large range in semimajor axis of the initial embryos, we assume that the 
surface density decreases as $r^{-3/2}$ \citep{wei97}. Of course these are 
tentative assumptions, since no data are available. Observations of circumstellar 
disks in the binary system L1551 by \cite{rod98} seem to suggest high density 
values: for a disk extending only for about 10 AU they derive masses of the 
order of 3 times the minimum mass of the solar nebula. However, in this case 
the evolutive state of the accretion disk and whether material is 
still falling onto the star is not known . 

In the simulations we used different values for the embryo's initial mass
ranging from 0.75 M$_{\rm L}$ ($\sim 0.01 {\rm M}_\oplus$) to 5.00 M$_{\rm L}$
($\sim 0.06 {\rm M}_\oplus$). The total mass in the embryos was about 75\% of 
that expected in a disk with $\sigma =$8 g\,cm$^{-2}$  since a 
consistent fraction possibly still resides in small planetesimals. Depending on the 
initial borders of the disk of planetary embryos, their total mass ranged from 
1.47 to 2.61 ${\rm M}_\oplus$. The maximum outer edge considered for the disk 
was 2.79, AU in agreement with the limit for orbital stability of \cite{hwi99}, 
while the inner border was set to 0.4 AU. The protoplanets radii, are calculated 
assuming a density of 3 g\,cm$^{-3}$ typical for planetary embryos in the 
terrestrial planet region \citep{cha98}. The initial eccentricities are 
randomly selected in the range $0 \le e \le 0.04$, the inclinations in the 
range $0^\circ \le i \le 0.6^\circ$. The remaining orbital elements are chosen 
randomly between $0^\circ$ and $360^\circ$. The initial separation between the 
planetary embryos, $\Delta$, is measured in mutual Hill radii, defined by the 
equation:
\begin{equation}
R_H = \Big(\frac{m_i + m_{i+1}}{3M_{\alpha\,{\rm Cen A}}}\Big)^{1/3}\Big(\frac{a_i + a_{i+1}}{2}\Big)
\end{equation}
where $i$ refers to the $i$-th body, specifically to the mass and to the 
orbital semimajor axis, $M_{\alpha\,{\rm Cen A}}$ is the mass of $\alpha$ Cen A.
We varied the value of $\Delta$ from $\sim$3 to 9 in the different simulations 
to test how this parameter influences the outcome.

The initial parameters are collected in Table \ref{tab2}.

\begin{table}
\center
\begin{tabular}{ccccccc}
   $n$  & $a_1$  & $a_n$  &  $m_i$  &  $m_T$  & $\Delta$\\
    &  AU    &  AU    &$M_{\rm moon}$&$M_\oplus$& $R_H$ \\
\hline
&&&&&\\
   100  &  0.55  &  1.64  &   1.25  &   1.53  &    3.65 \\
   100  &  0.55  &  1.93  &   1.50  &   1.83  &    3.95 \\
   100  &  0.55  &  2.24  &   1.75  &   2.14  &    4.19 \\
   100  &  0.55  &  2.58  &   2.00  &   2.44  &    3.11 \\
    77  &  0.75  &  2.10  &   1.75  &   1.65  &    4.00 \\
    64  &  0.55  &  1.68  &   2.00  &   1.56  &    4.00 \\
   128  &  0.50  &  1.82  &   1.10  &   2.02  &    3.50 \\
   175  &  0.50  &  2.37  &   1.05  &   2.57  &    3.13 \\
   112  &  0.50  &  1.74  &   1.20  &   1.94  &    3.75 \\
    36  &  0.50  &  2.36  &   5.00  &   2.57  &    9.00 \\
    39  &  0.50  &  2.27  &   4.50  &   2.47  &    8.50 \\
    41  &  0.50  &  2.14  &   4.00  &   2.35  &    8.00 \\
    44  &  0.50  &  1.98  &   3.50  &   2.18  &    7.50 \\
    45  &  0.50  &  2.10  &   3.50  &   2.10  &    7.00 \\
    53  &  0.50  &  1.75  &   3.00  &   1.94  &    7.00 \\
    52  &  0.50  &  1.92  &   3.00  &   1.91  &    6.50 \\
    52  &  0.50  &  1.61  &   2.50  &   1.59  &    6.00 \\
    82  &  0.50  &  2.11  &   2.00  &   2.32  &    5.00 \\
   150  &  0.55  &  1.93  &   1.00  &   1.83  &    3.00 \\
    95  &  0.55  &  1.84  &   1.50  &   1.74  &    4.00 \\
   117  &  0.55  &  2.65  &   1.75  &   2.50  &    4.00 \\
   137  &  0.75  &  2.58  &   1.25  &   2.01  &    3.00 \\
   149  &  0.50  &  2.43  &   1.25  &   2.61  &    3.50 \\
   131  &  0.50  &  1.33  &   0.75  &   1.47  &    2.95 \\
    77  &  0.75  &  2.10  &   1.75  &   1.65  &    4.00 \\
    64  &  0.55  &  1.68  &   2.00  &   1.56  &    4.00 \\
\end{tabular}
\caption{ List of the initial parameters for each of the
24 simulations. The parameters are:
$n$,  the number of bodies in each simulation;
$a_1$, the orbital semimajor axis of the inner body in AU;
$a_n$, the orbital semimajor axis of the outer body in AU; 
$m_i$, the mass of each body in Moon masses (Moon mass = 7.35$\cdot 10^{22}$ g);
$m_T$, the total mass of the disk of protoplanetary embryos in Earth masses (Earth mass = 5.976$\cdot 10^{24}$ g); and
$\Delta$, the separation between the bodies in mutual Hill radii.
}
\label{tab2}
\end{table}

\section{Typical evolution of a population of planetary embryos into planets}

\begin{figure}
\includegraphics[angle=0, width=8cm]{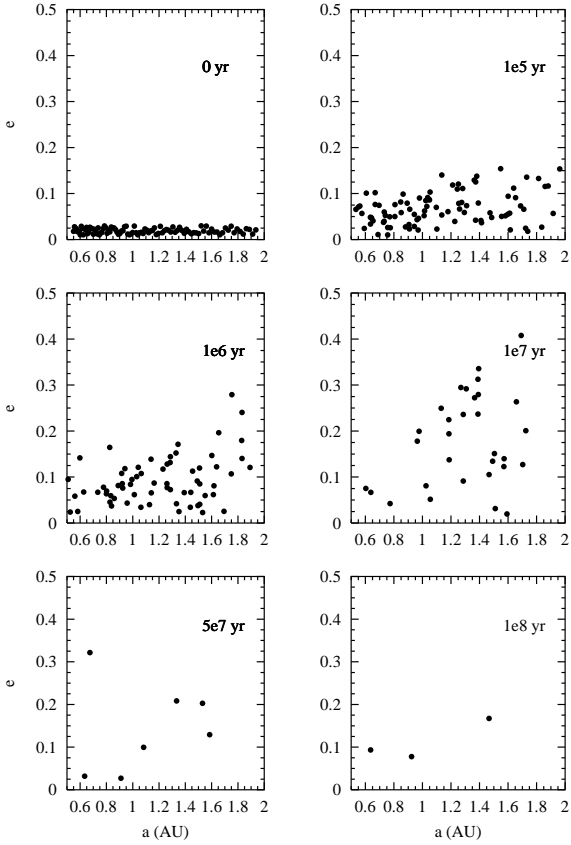}
\caption[]{
Snapshots at various times of the orbital semimajor axis $a$ versus the 
eccentricity $e$ for all the bodies in SIM2.
}
\label{f01}
\end{figure}

Here we present the evolution of a typical simulation of planetary formation 
in $\alpha$ Centauri taken from our set of 24 numerical cases. The initial 
parameters are given in Table \ref{tab2} (simulation number 2, hereinafter SIM2).
In Fig. \ref{f01} we show the time evolution of the semimajor axis versus 
eccentricity for all the bodies at different evolutionary times.
Already after 
10$^5$ years the orbits of planetary embryos are dynamically excited with 
eccentricities up to 0.1. At subsequent times the eccentricities grow even 
further for the combination of the companion star perturbations and mutual 
encounters and collisions. After 1 $\times 10^8$ years three planetary bodies 
are left: the two largest planets 
have a mass similar to the mass of Venus while the third one 
has two times the mass of Mars. The final orbital parameters are given in Table 
\ref{tab3}. 
\begin{table}
\begin{tabular}{lrrrrrrrr}
    &      mass & $a$      &$e$     &   $i$  &$\omega$&  $M$   &  $N$     \\
    &$M_\oplus$ &  AU      &        &$^\circ$&$^\circ$&$^\circ$&$^\circ$  \\
\hline
B02 &     0.550 &  0.637   & 0.094  &   1.7  &  209   &  115   &  333     \\
B31 &     0.715 &  0.924   & 0.078  &   1.5  &   35   &  186   &  218     \\
B36 &     0.202 &  1.467   & 0.167  &   3.4  &  327   &   67   &  325     \\
\end{tabular}
\caption{Orbital elements for the surviving planets in SIM2}
\label{tab3}
\end{table}

\begin{figure}
\includegraphics[angle=-90, width=8cm]{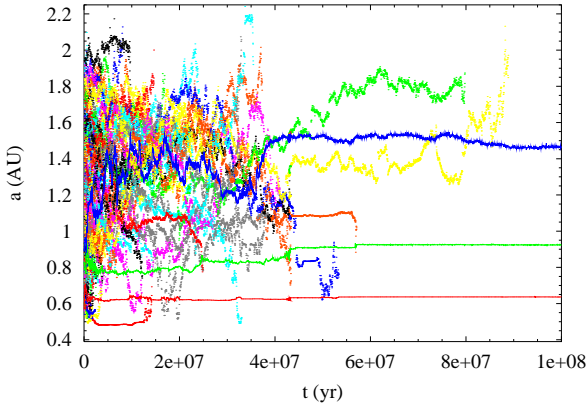}
\caption[]{
Orbital semimajor axis as function of time for all the bodies in simulation 2.
The surviving bodies are represented by continuous lines.
}
\label{f02}
\end{figure}

In Fig. \ref{f02} we illustrate the time evolution of the semimajor 
axes of each body in the simulation. 
After a brief chaotic evolution when all 
the bodies undergo repeated mutual close encounters, 
the number of bodies begins to 
decrease with time because of mutual collisions and consequent accumulation 
into 
larger bodies. Some of the initial embryos also impact on each of the two stars 
or are ejected out of the system. After 5$\times 10^7$ years most of the bodies 
have accreted into larger ones and only a few bodies are involved in the 
evolution of the system in the subsequent 5$\times 10^7$ years. The three 
surviving planets at the end of the simulation are in non--crossing orbits and 
form an apparently stable system.

The SIM2 is peculiar for its final dynamical configuration: the outer planet 
and the intermediate one are locked in a 2:1 mean motion resonance. This 
confirms that resonant configurations may be a common outcome of the planetary
formation process. Even two observed exoplanetary systems have their planets 
possibly locked in a 2:1 resonance, i.e. GJ876 \citep{mar01} and  HD82943 
\citep{goz01}. 

\begin{figure}
\includegraphics[angle=-90, width=8cm]{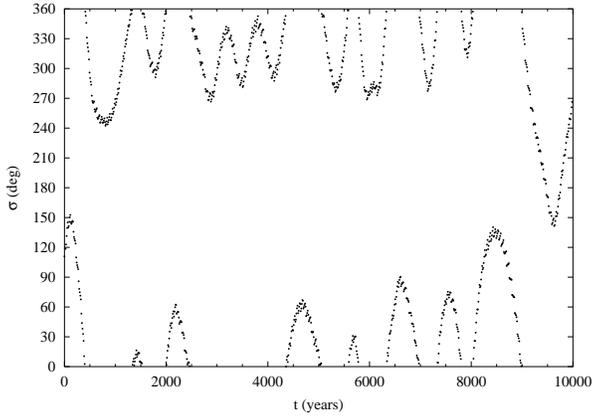}
\caption[]{
Critical argument $\sigma = 2 \lambda_2 - \lambda_1 - \tilde\omega_1$ 
of the 2:1 mean motion resonance detected between two planets at the end 
of SIM2.
$\lambda_1$ and $\tilde\omega_1$ are the mean longitude and perihelion 
longitude of the inner planet whose mass is 0.72 $M_{\oplus}$ and semimajor 
axis 0.92 AU;  $\lambda_2$ is the mean longitude of the outer planet whose 
mass is 0.2 $M_{\oplus}$ and semimajor axis 1.46 AU.
}
\label{f03}
\end{figure}

In  Fig. \ref{f03} we show the critical argument $\sigma = 2 \lambda_2 - \lambda_1 - \tilde\omega_1$ 
of the 2:1 resonance over a period of 1$\times 10^4$ years, where $\lambda_{1,2}$
are the mean longitudes of the inner and outer planets respectively and 
$\tilde\omega_1$ is the longitude of periapse of the inner planet. 
The resonant 
state between the two planets is however unstable and after 2.3 $\times 10^8$ 
years the outer planet begins to evolve chaotically, its eccentricity grows, 
and it is ejected out of the system leaving only the two major planets in 
orbit around $\alpha$ Cen A.

\section{Statistics for all the simulations}

\begin{figure}
\includegraphics[angle=-90, width=8cm]{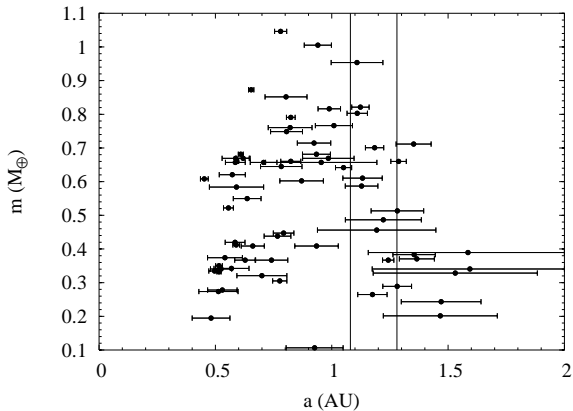}
\caption[]{
Distribution of semimajor axis vs. mass of all the terrestrial planets that 
formed in our simulations. The vertical lines are the edges of the habitable 
zone for $\alpha$ Cen A. The horizontal bars show the perihelion--aphelion 
distances. 
}
\label{f04}
\end{figure}

\begin{figure}
\includegraphics[angle=-90, width=8cm]{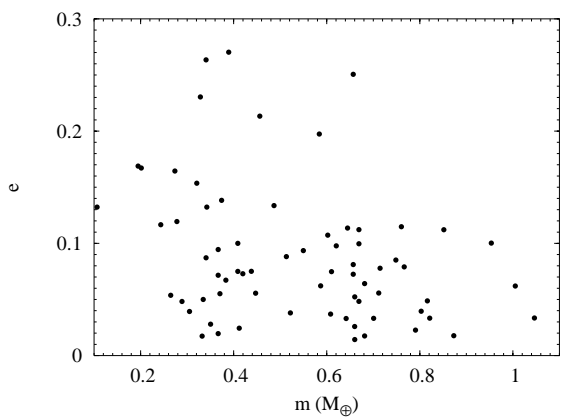}
\caption[]{
Eccentricity versus mass for all the planets formed at the end of our simulations.
}
\label{f05}
\end{figure}

In this section we combine the outcome of all the 24 simulations to produce 
statistical predictions on the planetary formation process in $\alpha$ 
Centauri. In Fig. \ref{f04} we plot the final mass of the planets that formed 
in all the simulations as a function of their semimajor axis.
The horizontal 
bars define the perihelion and aphelion distance for each body. The mass values 
range from 0.2 to $\sim$ 1  ${\rm M}_\oplus$ and some of them have their orbits within 
the habitable zone for $\alpha$ Cen A that, in the plot, is limited by two 
vertical bars. The maximum semimajor axis is $\sim$ 1.6 AU and bodies closer to this 
outer limit have larger values of eccentricity due to the gravitational 
perturbations of the companion star. The eccentricity vs. mass 
distribution shown in Fig. \ref{f05} confirms that some planets, 
in particular the inner ones, end up in almost circular orbits. 

Our Fig. \ref{f04} and Fig. \ref{f05} 
may be compared with  Fig. 11 and 12 of \cite{cha01} 
that illustrate similar results concerning the accretion of terrestrial 
planets 
in the solar system in presence of both Jupiter and Saturn
as perturbing planets. Since we used in our simulations the same value of 
$\sigma$ and similar values for the initial mass  and spacing of the embryos 
have been used, the comparison gives an  insight on the 
expected differences between the planetary formation process 
in the two systems. 
A similar triangular pattern is observed in the distribution of 
the final planetary mass vs. semimajor axis (our Fig. \ref{f04} and 
Fig. 11 of \cite{cha01}) but more massive planets may
form in the inner solar system, up to 1.5 $M_{\oplus}$, compared
to $\alpha$ 
Cen where the maximum mass value is 1 $M_{\oplus}$. 
In the simulations of \cite{cha01} low mass terrestrial planets 
may accumulate also beyond 1.5 AU, up to 2 AU while 
for $\alpha$ Cen we do not get planets beyond 1.6 AU.
In both systems the 
eccentricities are on average lower than 0.15 but we find 
a higher percentage of low eccentricity orbits in particular
for the inner planets. Moreover, 
in our Fig. \ref{f05} there is a more
clear trend towards higher values of eccentricity for less
massive bodies. 

A possible planetary system around $\alpha$ Cen A would then 
be very similar to the inner solar system with two or three terrestrial planets 
with size ranging from a Mars--like planet to an Earth--like planet.  
This is a somewhat surprising result taking into account the 
strong perturbations of the companion star.

\begin{figure}
\includegraphics[angle=-90, width=8cm]{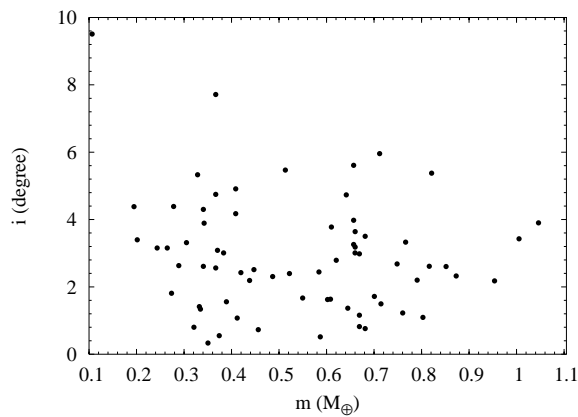}
\caption[]{
Mass of the final planets versus their orbital inclination.
}
\label{f06}
\end{figure}

The distribution of the final inclinations is shown in Fig. \ref{f06} as a 
function of the final mass of the planets.
High inclinations, of the order of 
some degrees, may be achieved in particular by the smaller bodies. For the 
larger planets the inclination is comparable to that of the terrestrial planets 
in our Solar System.

\begin{figure}
\includegraphics[angle=-90, width=8cm]{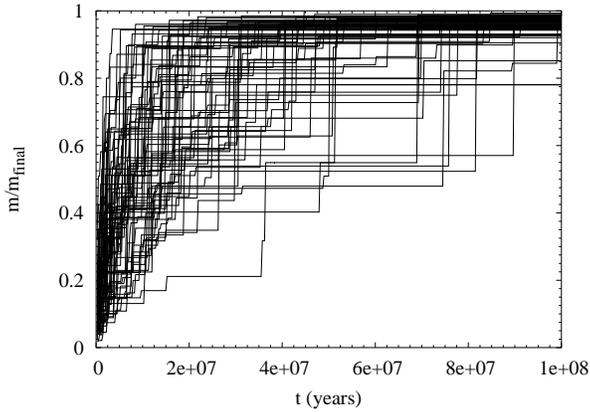}
\caption[]{Mass growth of the embryos that become planets vs. time. The 
value of mass
of each body is normalized to its final mass as a planet.}
\label{f07}
\end{figure}

To gain more insight on the evolution of planetary embryos into
planets, we have plotted in Fig. \ref{f07} 
the mass growth as a function of time of all the 
planets shown in Fig. \ref{f04}. 
For each body the mass
value is normalized to the final value. 
The majority of the virtual planets gain most of their mass within 
$4.0 \times 10^7$ years while in the subsequent years only 
about 10 \% of the final mass is accreted.  We can then set to 
$4.0 \times 10^7$ years the time-scale within which most of the
accretion occurs. 

\begin{figure}
\includegraphics[angle=-90, width=8cm]{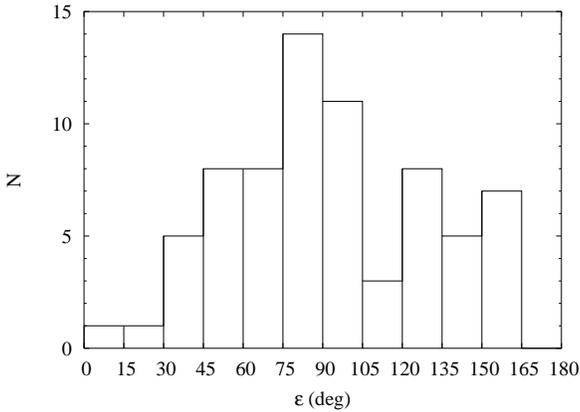}
\caption[]{Histogram of the obliquities of the final planets.}
\label{f08}
\end{figure}

In Fig. \ref{f08} we show the final values of 
the obliquity of the planets. 
The distribution is random with a
peak around 80$^\circ$.  Only a small fraction has 
an obliquity close to 0$^\circ$
that represents the state of our Solar
System with most of the planets having the spin axes perpendicular 
to the orbital plane.
Similar results on the obliquity were also obtained by \cite{cha01}.
One possible explanation is that in our simulations we have not considered
the effects of fragmentation and hence the possibility of formation 
of satellites
that might reduce the obliquity (as in the case of Earth and Moon).

At the end of each integration we have checked whether the
final planets are in some kind of mean motion resonance. 
In 6 simulations out of 24, two of the final
planets were close within 1\% in semimajor axis 
to a low order mean motion resonance (2:1 or 3:2). 
In one case, that described in Section 4 and termed SIM2, 
the two planets were in resonance as tested by the 
libration of the critical argument. 
This suggests that resonant configurations in planetary systems 
may not be rare. To give statistical robustness to this 
statement we should however estimate the width of the main 
resonances (possibly as a function of the eccentricity)
and compare it with the semimajor axis 
range available for stable orbital configurations of the planets. 
We could then state that 6 cases out of 24 is 
a statistical significant result and not just a random outcome. 
Such an investigation is however out of the scope of this paper.

\section{Ingestion of planetary material by the parent stars}

\begin{figure}
\includegraphics[angle=-90, width=8cm]{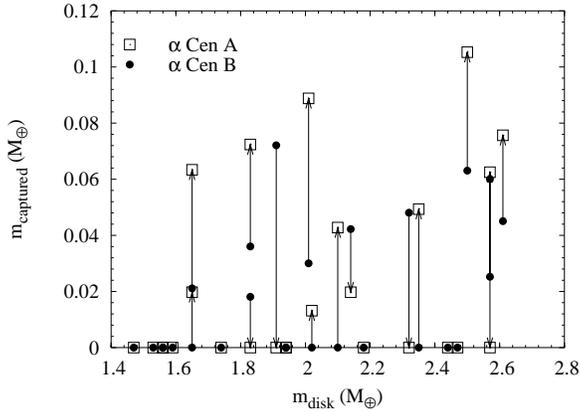}
\caption[]{
Amount of mass captured by the two stars during the growth of 
terrestrial planets around $\alpha$ Cen A.
}
\label{f09}
\end{figure}

During the process of planetary formation some embryos, because of the strong 
perturbations of the binary companion, 
impact on the primary or on the secondary 
star. In all our simulations we compute the fraction of the initial mass that 
ends up in either of the stars. In Fig. \ref{f09} we show the amount of mass 
plunging into $\alpha$ Cen A and B, respectively. 
On average, about 2\% of an 
Earth mass is engulfed by either of the stars, so the metallicity of the stars 
should not have been significantly altered in this final stage of the planetary 
accretion process. At least 0.5  ${\rm M}_\oplus$, according to \cite{mur01}, 
are needed to alter noticeably the metallicity of a main--sequence star. 
However, we started our simulations with a protoplanetary disk whose density 
was equal to that expected in the primordial solar nebula. More massive initial 
disks would lead to significantly higher amount of mass falling onto the two 
stars, possibly affecting the composition of the convective envelope. From 
Fig. \ref{f07} we also notice that there may be an asymmetry between the amount 
of material plunging onto the two stars and, as a consequence, there might be 
a difference in the metallicity of a close binary system if planetary accretion 
occurred only around one of the components.  

We have to point out that the previous results on the mass infall of material 
on the stars must be taken as indicative in a statistical sense. Numerical 
integrations of highly eccentric protoplanetary trajectories may be not very 
accurate when the bodies are far from the primary body. 
When a protoplanet in a high eccentric orbit 
travels around the aphelion and then closer to 
the companion star, the orbital integration may accumulate a 
significant integration error. However, being the orbits 
easily chaotic, the outcomes of our integration in this case
may be taken as predictive only in terms of probability of impact onto 
the two stars among the escaping bodies.

\section{Conclusions}

Planetary formation seems to be possible in the Alpha Centauri system
and, in general, in close binary systems. \cite{msc00} 
showed that planetesimals can accrete into planetary embryos; in 
this paper we demonstrate that planetary embryos can grow into
planets in about 50 Myr. These terrestrial--size planets form 
within the region of planetary stability found by  \cite{who97}
and are possibly stable over the age of the system. 

Assuming that the initial mass in the protoplanetary disk 
around the main component of the $\alpha$ Cen system
was comparable to that of the minimum solar nebula, planets with 
the size of Mars and Venus are typically formed with 
low orbital eccentricity. Some of them lay within 
the habitable zone of the star and, possibly, could harbor life. 
 
During the accumulation process, some of the planetary embryos are
engulfed by the stars in different amounts. However, only if
the protoplanetary disk was very massive, at least five times more
than the minimum mass solar nebula, we might observe a difference
in the metallicity of the star related to the ingestion process. 

Orbital resonances may occur between the planets. In a significant
fraction of our numerical simulations we found an orbital period 
close to a low order commensurability ratio. In one case we also found 
a real 2:1 resonance between two planets, with the critical 
argument librating around 0. This system was stable over 
200 Myr after which the outer planet entered a chaotic state and 
was ejected out of the system.

\end{document}